\documentclass[a4paper,aps,prb,twocolumn,groupedaddress]{revtex4}

\usepackage{graphicx}
\usepackage{dcolumn}
\usepackage{bm}
\usepackage{amsmath}

\begin{document}

\title{Statistical Electron Excitation in a DQD Induced by Two Independent Quantum Point Contacts  }

\author{U. Gasser}
\author{S. Gustavsson}%
\author{B. K\"{u}ng}%
\author{K. Ensslin}%
\author{T. Ihn}%
\affiliation{%
Solid State Physics Laboratory, ETH Zurich, CH-8093 Zurich,
Switzerland
}%
\author{D.C. Driscoll}%
\author{A.C. Gossard}
\affiliation{ Materials Department, University of California, Santa
Barbara, California 93106, USA}%

\date{\today}

\begin{abstract}
We investigate experimentally the influence of current flow through
two independent quantum point contacts to a nearby double quantum
dot realized in a GaAs-AlGaAs heterostructure. The observed current
through the double quantum dot can be explained in terms of coupling
to a bosonic bath. The temperature of the bath depends on the power
generated by the current flow through the quantum point contact. We
identify the dominant absorption and emission mechanisms in a double
quantum dot as an interaction with acoustic phonons. The experiment
excludes coupling of a double quantum dot to shot-noise generated by
quantum point contact as the dominant mechanism.

\end{abstract}

\pacs{Valid PACS appear here}
\maketitle


\section{Introduction}\label{sec:Introduction}

Electronic transport through semiconductor double quantum dots
(DQDs) has been intensively explored for nearly two
decades.~\cite{Reed,vanderWiel:1,Hanson} The interplay between a
double quantum dot and its environment was investigated in detail in
previous
works~\cite{Fujisawa:36,Fujisawa:21:247,Oosterkamp,Oosterkamp:Nature,vanderWiel:272,Blick:PAT,Switkes}
using microwave spectroscopy. Irradiating double quantum dots with
microwaves results in photon assisted tunneling
(PAT).~\cite{Platero} The integration of a quantum point contact
(QPC) in the vicinity of a single quantum dot allowed charge
detection,~\cite{Field:70.1311} which was later implemented in
double quantum dot systems.~\cite{Elzerman2003,DiCarlo}

The novel application of a quantum point contact as a source of
energy to drive inter-dot electronic transitions in a double quantum
dot was recently
realized.~\cite{Khrapai:97.17,Khrapai:2008.1,Khrapai:2008.2} These
experiments were explained in terms of acoustic phonon based energy
transfer between the QPC and the DQD circuits. The combination of a
capacitatively coupled DQD-QPC system with time resolved charge
detection resulted in a frequency-selective detector for microwave
radiation. It allows to detect single photons emitted by the QPC and
absorbed by the DQD.~\cite{Gustavsson:99}

Understanding the back-action of a charge sensor on a DQD is
important for future possible applications in quantum information
processing.~\cite{Fujisawa:2006:Single} The possible dominant
mechanisms that lead to QPC-induced inter-dot electronic transitions
include electron scattering with photons~\cite{Gustavsson:99} and
acoustic phonons~\cite{Fujisawa:Science98} or
shot-noise~\cite{Onac,Zakka} depending on the parameter regime
investigated.

In this paper we study back-action of the current flow through the
QPC detector on a serial double dot. The double dot is tuned to an
asymmetric regime, where one dot is strongly coupled to the source
lead, whereas the second dot is more weakly coupled to the drain
lead. Two independent QPCs can be simultaneously used for driving
the transitions in the DQD. We observe a non-additive effect of both
QPCs accompanied by the saturation of the current across the double
quantum dot for large QPC currents. We explain the measured data in
the framework of interaction of electrons with acoustic phonons. We
relate the power emitted by the QPC to the temperature of the
phononic bath. The experiment excludes the possibility of shot-noise
being the source of inter-dot transitions.

This paper is organized as follows. In Sec.~\ref{sec:Sample} we
describe the fabrication of the sample, its electrostatic
characterization and functionality. In Sec.~\ref{sec:Experimental}
we present a detailed description of the working regime of a DQD and
QPCs, followed by the results of our measurements of current through
a DQD using one and two QPCs. We discuss the possible interaction
mechanisms in Sec.~\ref{sec:Mechanisms}. In Sec.~\ref{sec:Model} we
introduce a model based on electron-phonon interaction and in
Sec.~\ref{sec:Results} we interpret the measured data. Section
\ref{sec:Conclusions} contains the conclusions.

\section{Sample and characterization}\label{sec:Sample}

\begin{figure}[h!]
\includegraphics[width=86mm]{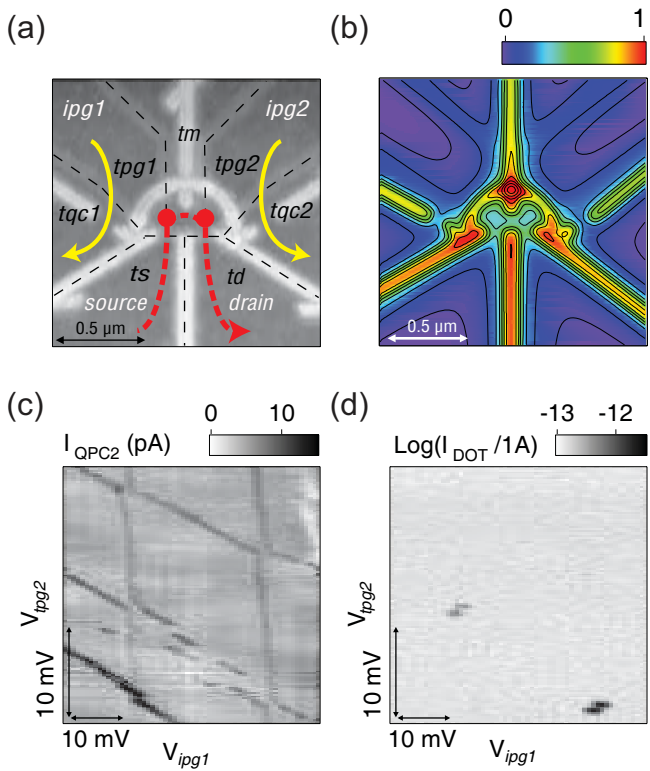}
\caption{\label{fig:intro}(Color online) (a) SFM image of the
sample. In-plane gates are defined by the thick white lines,
titanium oxide lines are indicated by dashed lines. Top gates are
labeled with black letters. The QDs are indicated with grey (red
online) circles. The white (yellow online) arrows indicate the
positive direction of the current in the QPCs. (b) Calculated
electrostatic potential in the 2DEG generated by the oxide lines and
the top gates. (c) Modulated current component proportional to the
transconductance measured through QPC2 for V$_{\mathrm{QPC2}}$=0.5
mV. An AC voltage of 1.5 mV with a frequency of 34 Hz was applied to
the in-plane gate \emph{ipg1} and detected in the QPC2 circuit using
lock-in techniques. The honeycomb pattern and four pairs of triple
points are visible. (d) Simultaneously measured double dot DC
current. The source-drain voltage was V$_{SD}$=60~$\mu$V.}
\end{figure}

The sample shown in Fig.~\ref{fig:intro}(a) is based on a
GaAs/Al$_{\mathrm{0.3}}$Ga$_{\mathrm{0.7}}$As heterostructure with a
two-dimensional electron gas (2DEG) 34 nm below the surface. It was
fabricated by double layer local oxidation with a scanning force
microscope (SFM).~\cite{sigrist:3558} The 2DEG is depleted below the
oxide lines written on the GaAs surface~\cite{Fuhrer} [white lines
in Fig.~\ref{fig:intro}(a)]. A 4 nm titanium film was evaporated and
patterned by local oxidation to create mutually isolated top gates
[indicated by the dashed lines in Fig.~\ref{fig:intro}(a)].

The confinement potential produced by the top gates and the oxide
lines is shown in the contour plot in Fig.~\ref{fig:intro}(b). It
was calculated assuming a pinned surface model~\cite{Davies:1} using
the lithographic sizes of the gates measured after the sample was
fabricated. It shows an approximately circular symmetry for the
dots, with the left quantum dot being slightly larger than the right
one. The color scale is in arbitrary units.\cite{potential}

The structure presented in Fig.~\ref{fig:intro}(a) consists of three
electronic circuits. The first one is formed by two quantum dots
connected in series [marked by the grey (red online) circles] and
connected to source and drain. A negative DQD current corresponds to
electrons moving from source to drain. Each of the other two
circuits contains a quantum point contact [white (yellow online)
solid arrows]. A negative QPC current means electrons traveling
through the QPC in the direction of the arrows.

To sum up, our structure consists of two barriers defining quantum
point contacts, two quantum dots, two barriers determining the
coupling of the quantum dots to the source and drain and the barrier
that determines the coupling between the quantum dots. In total,
this gives seven degrees of freedom and there are seven independent
top-gates used to tune these barriers and the quantum dots. The top
gates (\emph{tpg1} and \emph{tpg2}) are used to tune the DQD into a
suitable regime. The top gates \emph{tqc1} and \emph{tqc2} can tune
the transmission of QPC1 and QPC2, respectively. The middle top gate
\emph{tm} controls the coupling between the two dots allowing to
change smoothly from the single dot regime (large dot spread over
the area covered by the two red circles) to a weakly-coupled double
dot. The gates \emph{ts} and \emph{td} are used to tune the coupling
of the DQD to source and drain.

The potential on both sides of QPC1 (QPC2) can be lifted with
respect to the measurement ground, creating a mutual gating effect
between DQD and QPC1 (QPC2). These in-plane gates (\emph{ipg1} for
QPC1 and \emph{ipg2} for QPC2) control the number of electrons on
the DQD.

Due to the presence of the metallic top gates, the electrostatic
interaction between electrons in the quantum dots and the QPCs is
weakened by screening compared to semiconductor-only quantum
circuits.~\cite{gustavsson:152101} The large distance between the
QPC and the double dot (lithographic distance 450 nm) further
reduces the sensitivity of the QPC for detecting electrons passing
through the DQD.

Figures ~\ref{fig:intro}(c) and (d) demonstrate the operation of
QPC2 as a charge detector.~\cite{Field:70.1311} For both QPCs, the
one-dimensional subband spacing is larger than 3.5 mV as estimated
from finite bias measurements.  In order to use QPC2 (QPC1) as a
charge read-out, its conductance was tuned to $e^{2}/h$. A constant
voltage of 0.5 mV was applied between the source and drain leads of
the QPC, and the current was measured. An AC voltage of 1.5 mV
applied at 34 Hz to the opposite in-plane gate \emph{ipg1}
(\emph{ipg2}) modulated the current through QPC2 (QPC1). This
modulated signal which is proportional to the transconductance was
detected with lock-in techniques. The measurements were performed in
a dilution refrigerator at a base temperature of 70 mK.

The resulting stability diagram of the DQD detected with QPC2 is
shown in Fig.~\ref{fig:intro}(c). The boundaries between regions of
different ground state charge configurations of the DQD are clearly
visible. In this measurement, \emph{tpg2} is used to change the
number of electrons in the right dot and \emph{ipg1} to change the
number of electrons in the left dot. A few charge rearrangements in
the lower half of the honeycomb induced by the metallic top gate
\emph{tpg1} are present. In general, we find that the top gate
sweeps lead to significantly more charge rearrangements than sweeps
of the in-plane gates. Change of the \emph{ipg2} potential combined
with simultaneous charge detection would result in strong detuning
of the charge sensor. This would shift the operating point far away
from the sensitive regime. It can be avoided by using \emph{tpg2}
because it has a weaker lever arm on the QPC. The thick line in the
bottom-left corner of the plot corresponds to a resonance in QPC2.

In Fig.~\ref{fig:intro}(d) the corresponding DC current through the
DQD is plotted. It was measured simultaneously with the QPC signal
presented in the previous paragraph. The source-drain voltage
applied to the DQD is 60~$\mu$V. Only two pairs of triple points are
visible. Similar sets of data can be obtained using QPC1 as the
detector.

\begin{figure}[h!]
\includegraphics[width=86mm]{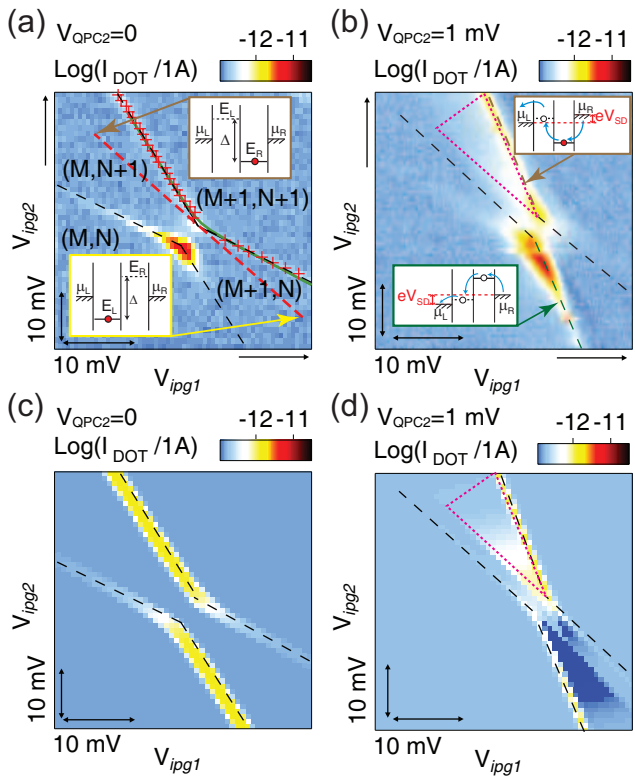}
\caption{\label{fig:SD}(Color online) (a) Double dot current as a
function of the gate voltages $V_{\emph{ipg1}}$ and
$V_{\emph{ipg2}}$. It represents the charge stability diagram of the
DQD. The dashed lines outline the edges of the honeycomb
cells.~\cite{vanderWiel:1} M and N refer to the number of electrons
in the left and the right dot, respectively. The red line denotes
the detuning axis, with zero detuning occurring at the triple point.
The red crosses correspond to the positions of the current maxima.
The green line depicts the calculated charge configuration diagram
with the tunneling coupling $t=50~\mu$eV. It was obtained assuming
the constant interaction model with a finite tunnel
coupling.~\cite{vanderWiel:1}
 The data was taken with $V_{\mathrm{SD}}=100~\mu$V applied symmetrically across the DQD, lifting the
 chemical potential in the source lead $\mu_{\mathrm{L}}$ up and lowering the energy of the
 drain lead $\mu_{\mathrm{R}}$. The insets schematically show the level alignment at
 points indicated by the arrows.
(b) The same region as in (a) but with a 1 mV DC bias voltage
applied across QPC1 and $60~\mu$V applied symmetrically across the
DQD. The triangle indicates the region with the induced DQD current.
The insets show schematic diagrams of the electrochemical potentials
in the left and the right quantum dot along the selected honeycomb
boundaries. (c) Calculated current through a double dot as a
function of gate voltages. The calculation refers to the measurement
presented in panel (a), where no bias voltage was applied to the
QPC. The details of the calculation are discussed in the text. (d)
Calculated stability diagram corresponding to the situations in (b),
where a voltage of 1 mV was applied across the QPC2. The dark blue
color in (d) was used to mark the region with a negative current.
The comparison between (b) and (d) is presented in the text.}
\end{figure}

\section{Experimental data}\label{sec:Experimental}

In the following, we concentrate on a single pair of triple points
where the DQD showed moderate coupling. Figure~\ref{fig:SD}(a) shows
the DC DQD current (I$_{\mathrm{DOT}}$) for 100~$\mu$eV source-drain
bias applied across the DQD. The inter-dot mutual capacitance
$C_{m}$ estimated from finite bias measurements and from the
stability diagram assuming the constant interaction
model~\cite{vanderWiel:1} is 8.8 aF, whereas the total capacitance
of the left dot is $C_{1}=86$ aF and for the right dot, $C_{2}=76$
aF. Each dot contains approximately 15 electrons and the charging
energies are about 2 meV.

The thin dashed lines in Fig.~\ref{fig:SD}(a) indicate the
boundaries of the honeycomb pattern and the numbers in brackets
(M,N) denote the charge population of the left and the right quantum
dot, respectively. Here, the left dot is strongly coupled to the
source lead, whereas the right dot is weakly coupled to the drain
reservoir. During the measurement, both QPCs were kept at zero bias.

The detuning marked by the dashed grey (red online) line in
Fig.~\ref{fig:SD}(a) is obtained from the capacitance
model~\cite{vanderWiel:1} and expressed by the equation
$\delta=E_{\mathrm{L}}-E_{\mathrm{R}}$ such that the total energy of
the DQD, $E_{\mathrm{tot}}=E_L+E_R$ remains constant. The energies
$E_L$ and $E_R$ are the single-particle energies in the left and the
right quantum dot, respectively. Converting the energies to gate
voltages gives:
$\delta=(\alpha_{\emph{ipg1,R}}-\alpha_{\emph{ipg1,L}})\Delta
V_{\emph{ipg1}}+(\alpha_{\emph{ipg2,R}}-\alpha_{\emph{ipg2,L}})\Delta
V_{\emph{ipg2}}$. The lever arms $\alpha_{\emph{ipg1,j}}$ and
$\alpha_{\emph{ipg2,j}}$ are the lever arms of the in-plane gates
\emph{ipg2} and \emph{ipg2} on the left ($j$=$L$) or the right
($j$=$R$) dot, respectively. The voltages $V_{\emph{ipg1}}$ and
$V_{\emph{ipg2}}$ are the voltages applied to the gates \emph{ipg1}
and \emph{ipg2}. The lever arms are extracted from measurements at
finite bias and from the charge stability diagram of the DQD. The
obtained values are: $\alpha_{\emph{ipg1,R}}=0.048$,
$\alpha_{\emph{ipg1,L}}=0.021$, $\alpha_{\emph{ipg2,R}}=0.03$ and
$\alpha_{\emph{ipg2,L}}=0.04$. We take zero detuning to occur at the
triple point. According to the definition above, detuning is
positive (negative) in the upper-left (lower right) part of
Fig.~\ref{fig:SD}(a). Two representative energy diagrams are shown
in the insets.

In Fig.~\ref{fig:SD}(b) the DQD current was measured in the same
parameter range at a QPC2 bias voltage of 1 mV. The bias voltage
across DQD was set to 60~$\mu$eV, i.e., smaller than the bias
voltage applied in Fig.~\ref{fig:SD}(a). Despite that, the current
is strongly enhanced along the boundaries (M,N)$\rightarrow$(M+1,N)
and (M,N+1)$\rightarrow$(M+1,N+1), corresponding to adding an
electron to the left dot. The enhancement of the current along the
honeycomb boundaries is induced by driving a current through QPC2.
Another visible feature induced by biasing QPC2 is the finite DQD
current in the triangle-shaped area indicated in
Fig.~\ref{fig:SD}(b) that is normally forbidden by Coulomb blockade.

\begin{figure}[h]
\includegraphics[width=86mm]{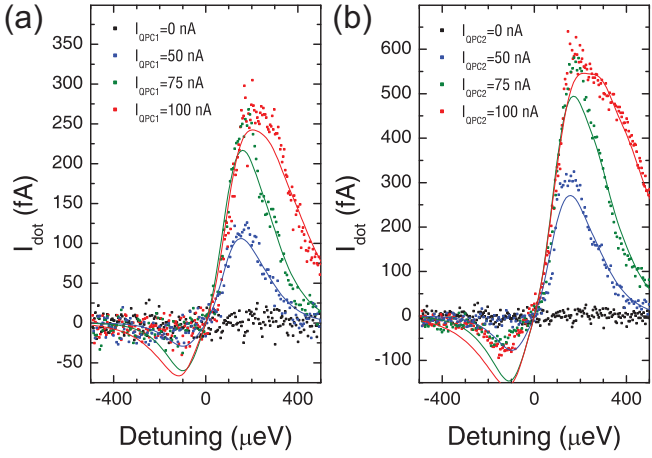}
\caption{\label{fig:detuning}Current through the DQD along the
detuning line: (a) for different QPC1 currents and (b) different
QPC2 currents. Fits are plotted with solid lines (see text).}
\end{figure}

The following measurements were carried out with the QPCs tuned to
their first conductance plateau. The overall experimental results do
not depend on this operation point of the QPC.

To investigate the influence of the QPC currents on the DQD in the
triangular region, we tuned the levels in the dot along the detuning
line depicted as the solid red line in Fig.~\ref{fig:SD}(a).
Fig.~\ref{fig:detuning} shows the dot current versus detuning. The
black data points in Fig.~\ref{fig:detuning}(a) were taken with zero
bias applied to the DQD as well as to QPC1 and QPC2. No measurable
current above the noise level is detected. When a DC current of 50
nA is driven through QPC1, an asymmetric peak with a maximum of
about 125 fA along the detuning line is observed (blue points in
Fig.~\ref{fig:detuning}(a)). This effect is strongly enhanced if the
current through QPC1 is further increased to 75 nA (green points)
and 100 nA (red points). All traces cross zero at the triple point
(zero detuning).

A similar, but significantly more pronounced effect is observed if
QPC2 is driven, as shown in Fig.~\ref{fig:detuning}(b). Moreover,
for negative detuning a small negative DQD current is observed. QPC2
is more sensitive as a charge readout and it has a stronger effect
on the DQD. Therefore, we conclude that QPC2 is more strongly
coupled to the DQD than QPC1.

\begin{figure}[h!]
\includegraphics[width=86mm]{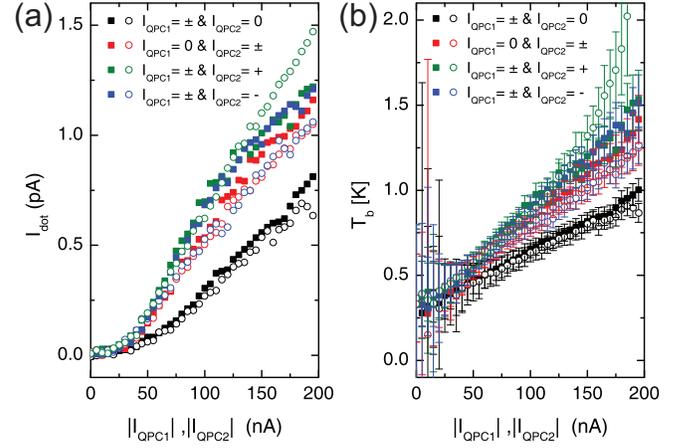}
\caption{\label{fig:addition}(a) Current through the DQD at fixed
detuning of 200~$\mu$eV as a function of QPC1 and QPC2 current. The
open and closed symbols correspond to the "$+$" and "$-$" sign in
the corresponding expression, respectively. (b) Temperature of the
bosonic bath as a function of the QPC1 and QPC2 current calculated
from (a) (see text).}
\end{figure}

We have chosen one point on the detuning line corresponding to
$\delta$=200~$\mu$eV and swept the QPC1 and QPC2 currents. The
results of this measurement are shown in Fig.~\ref{fig:addition}(a).
The black (red) filled squares correspond to positive currents
through QPC1 (QPC2) swept from 0 to 200 nA. The empty black (red)
circles are the traces recorded while the QPC1 (QPC2) current was
swept from 0 to $-200$ nA. The QPC induced DQD current is a little
larger in the case when the QPCs are driven with positive current.
This polarity dependence is significant and we can exclude that it
is due to a gating effect. As mentioned before, QPC2 is more
strongly coupled to the DQD than QPC1. When the QPC current is swept
in a positive direction (filled symbols in Fig.~\ref{fig:addition})
the DQD current starts to level off (the inflection points for are
lying between 100 and 150 nA in the QPC current axis), whereas for
negative QPC current directions this effect is not clearly visible.

Another unexpected feature is observed on the green traces. The
filled (empty) green squares correspond to the QPC1 current being
swept from 0 to 200 ($-200$)~nA while the QPC2 current is
simultaneously swept from 0 to 200~nA. In a simple picture, we would
expect that the effects of QPC1 and QPC2 are independent and they
add up, but the measurement contradicts this expectation. Due to the
action of both QPCs the DQD current is slightly larger than in the
case when only QPC2 is used. In addition, there is an unexpected
polarity dependence with a maximum DQD current for QPC1 being swept
in negative and QPC2 in positive direction. The remaining blue
filled squares (empty circles) in Fig.~\ref{fig:addition}(a) were
obtained by driving a positive (negative) current through QPC1 and a
negative current through QPC2.
\begin{figure}[h]
\includegraphics[width=86mm]{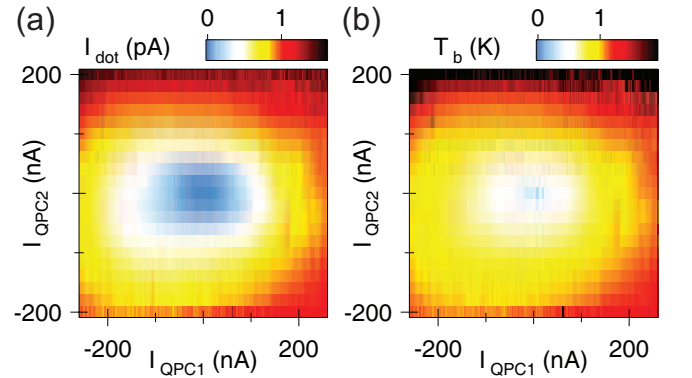}
\caption{\label{fig:fits}(Color online) (a) The DQD current as a
function of the QPC1 and QPC2 currents. The data was taken at a
fixed detuning $\delta=200~\mu$eV. (b) Calculated temperature of the
bosonic bath $T_{\mathrm{b}}$.}
\end{figure}

The polarity effect in the DQD current is also present in
Fig.~\ref{fig:fits}(a), where the dot current was plotted versus
QPC1 and QPC2 currents at fixed detuning $\delta$=200~$\mu$eV. In
this measurement, the lack of additivity of the effects induced by
the QPCs is even more visible.

\section{Discussion of possible mechanisms}\label{sec:Mechanisms}

\begin{figure}[h]
\includegraphics[width=86mm]{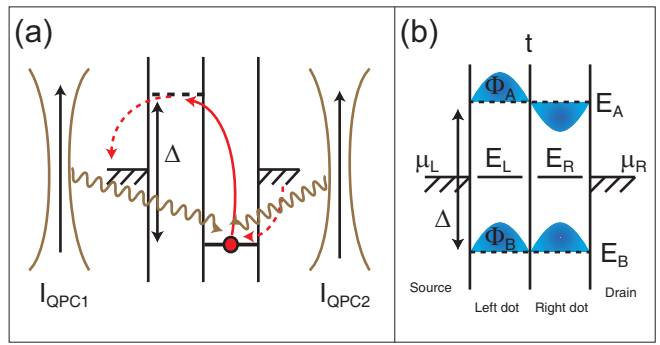}
\caption{\label{fig:scheme}(Color online) (a) Driving current
through the QPC induces a current through the DQD. An electron
absorbs an energy quantum $\Delta$ emitted from the left or the
right QPC and is excited to the excited state. It leaves the dot via
the left barrier and another electron can tunnel from the right into
the right dot. (b) Schematic energy level diagram for the DQD. The
wave functions of the bonding ($\Phi_{\mathrm{B}}$) and antibonding
($\Phi_{\mathrm{A}}$) states are shown. The electrochemical
potential of the left (right) lead is $\mu_{\mathrm{L}}$
($\mu_{\mathrm{R}}$). }
\end{figure}

A mechanism which can induce the current flow through the double dot
along the detuning line is presented in Fig.~\ref{fig:scheme}(a).
The driving current through QPC1 or/and QPC2 is thought to lead to
an emission of energy, which can be absorbed by the electron in the
right dot. If the provided energy matches the energy difference
$\Delta$, the electron can be excited from the right to the left
dot. If the electron leaves the DQD through the left lead and the
next electron tunnels into the right dot through the right lead,
then the cycle closes and there is a measurable current flowing
through the double dot.

An additional enhancement of the DQD current along the honeycomb
boundaries as observed in Fig.~\ref{fig:SD}(b), induced by driving a
current through a QPC, can be explained in a similar way. In the
situation shown in the upper inset the electron trapped in the right
quantum dot can absorb energy emitted by the QPC2, tunnel into the
left dot and leave the DQD system via the left lead. The cycle
closes when the next electron tunnels into the right dot from the
right lead. This QPC2 induced process gives an additional
contribution to the DQD current. This effect is more pronounced in
the vicinity of the triple point where energy difference between the
levels $E_L$ and $E_R$ in the left and the right dots are small. The
lower inset of Fig.~\ref{fig:SD}(b) shows the analogous diagram for
the situation when the level in the right quantum dot lies above the
Fermi energy of the leads. Again, the QPC2 induced process causes
the electrons to move from the right into the left contact.

The possible mechanisms of the pumping effect are coupling to
acoustic or optical phonons, plasmons, photons, shot-noise or
thermopower effect. Scattering with optical phonons is strongly
suppressed as long as the relevant energy scales are smaller than
the optical phonon energy.~\cite{Bockelmann} Coupling to plasmons
can be ruled out as well.~\cite{Fujisawa:Science98} We can also
exclude the shot-noise as a source of the energy, because during the
experiment both QPCs were tuned to their first plateau. Measurements
performed at 0.5$G_0$ and 1.5$G_0$ ($G_0=2e^2/h$) showed a
qualitatively and quantitatively similar behavior. This is in
contrast to previously measured data~\cite{Onac,Khrapai:97.17} where
no DQD current was observed in the plateau regions. Coupling to
acoustic phonons is the most likely mechanism of inducing the
current in the DQD. Further below in this paper, we discuss the data
in the light of phonon coupling and a related thermopower effect.

The questions arising from the data presented above are the
following: is the strong difference of the peak heights on the
positive and negative side of the detuning (Fig.~\ref{fig:detuning})
due to the asymmetry in the DQD coupling to the leads? Why do the
effects of the QPCs not add up? What is the reason for the
saturation of the DQD current observed in
Fig.~\ref{fig:addition}(a)? What is the mechanism of the energy
transfer from the QPCs to the DQD? Can it explain the polarity
dependence? In the next section we present a model that attempts to
answer most of these questions.

\section{The model}\label{sec:Model}

In the following we derive a scenario, which explains the pumping
effect based on electron-phonon interaction. First, we introduce the
two-level system describing the DQD. Then, we consider all possible
transitions between different energy states of the DQD and express
them in terms of tunneling rates. Subsequently, we derive the energy
dependence of the tunneling rates. The intra-dot transitions are
calculated in a framework of electron-phonon interaction. Next, we
set up a master equation and obtain the complete expression for the
DQD current as a function of detuning and the temperature of the
phonon bath.

Close to a pair of triple points, a double quantum dot can be
regarded as a two-level system,\cite{Blick:molecular,vanderWiel:1}
whose bonding (ground) and antibonding (excited) states $E_{B}$ and
$E_{A}$ are separated by an energy $\Delta=\sqrt{\delta^2+4t^{2}}$
as shown in Fig.~\ref{fig:scheme}(b),\cite{Blick:molecular} where
$t$ is tunneling coupling between the dots. The corresponding
eigenvectors for bonding and antibonding states are
$|\Psi_{\mathrm{B}}\rangle$ and $|\Psi_{\mathrm{A}}\rangle$. The
components of the bonding and antibonding eigenstates in the basis
of $|\phi_{\mathrm{L}}\rangle$ and $|\phi_{\mathrm{R}}\rangle$, the
wave functions in the left ($L$) and the right ($R$) dot, are
$c_{i,j}=\langle\phi_{\mathrm{i}}|\Psi_{\mathrm{j}}\rangle$, where
$i=\mathrm{L,R}$ and $j=\mathrm{B},\mathrm{A}$.

As in Sec.~\ref{sec:Experimental} the detuning is defined as
$\delta=E_{L}-E_{R}$ and the total energy is
$E_{\mathrm{tot}}=E_{\mathrm{L}}+E_{\mathrm{R}}$. We assume, that
the number of the electrons in the quantum dots is fixed and its
ground state energy is $E_{\mathrm{GS0}}=0$. If we add one extra
electron called an excess electron the ground state energy will be
$E_{\mathrm{GS1}}=\frac{1}{2}(E_{\mathrm{tot}}-\Delta)$, while for
two excess electrons, the ground state energy is
$E_{\mathrm{GS2}}=\frac{1}{2}(E_{\mathrm{tot}}+\Delta)+e^{2}/C_{m}$.
The corresponding electrochemical potentials are
$\mu_{1}=E_{\mathrm{GS1}}$ and
$\mu_{2}=E_{\mathrm{GS2}}-E_{\mathrm{GS1}}$.

\begin{figure}[h]
\includegraphics[width=86mm]{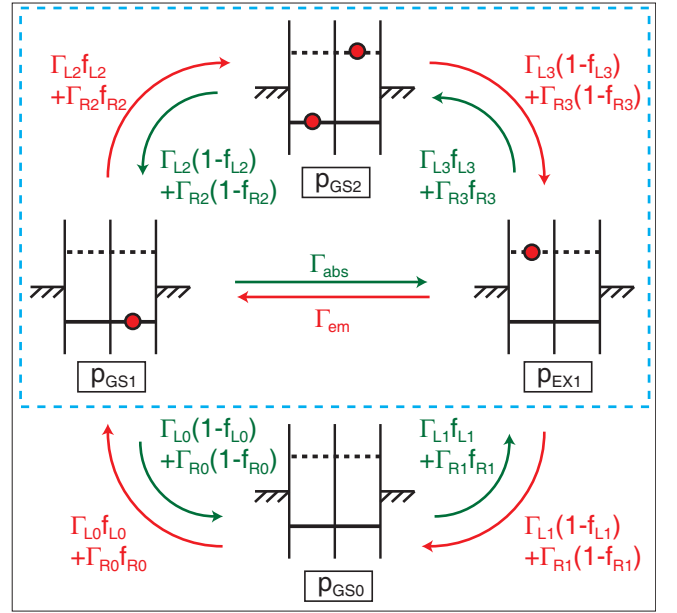}
\caption{\label{fig:model}(Color online) Energy level diagrams
presenting four possible charge states of the DQD in the vicinity of
a pair of triple points. The state $\mathrm{GS0}$ corresponds to the
situation with no excess electron present in the double dot. The
states $\mathrm{GS1}$ and $\mathrm{EX1}$ are the ground and excited
state of the double dot in the case of one excess electron,
respectively. The fourth state $\mathrm{GS2}$ is a two-excess
electron ground state. The occupation probabilities of these states
are labeled as $p_\mathrm{GS0}$, $p_{\mathrm{GS1}}$,
$p_{\mathrm{EX1}}$ and $p_{\mathrm{GS2}}$. The transitions between
these states are determined by tunneling rates $\Gamma_{i,j}$ and
Fermi function $f_{i,j}$. The index $i=(\mathrm{L,R})$ correspond to
the transition occurring through the left $i=(\mathrm{L})$ and
through the right $i=(\mathrm{R})$ lead. The second index
$j=0,1,2,3$ labels the transition of the electron in the DQD.}
\end{figure}

In the vicinity of a pair of triple points a double quantum dot can
have one out of four different charge states.  These different
charge configurations are presented in Fig.~\ref{fig:model}. The
"empty" state corresponds to a situation where there is no excess
electron present in a dot and the occupation probability of this
state is $p_{\mathrm{GS0}}$. The index GS0 denotes the zero-electron
ground state. In addition, one excess electron may occupy the
bonding (ground) state with probability $p_{\mathrm{GS1}}$ or the
antibonding (excited) state  with probability $p_{\mathrm{EX1}}$.
The last possible charge configuration is when there are two excess
electrons in a double quantum dot (two-electron ground state) with
occupation probability $p_{\mathrm{GS2}}$.

The transitions between these states are determined by the tunneling
rates $\Gamma_{i,j}$ and the thermal broadening of the Fermi
function $f_{i,j}$ in the leads. The index $i$=L,R denotes the left
(L) or the right (R) barrier through which the electron tunnels and
the index $j$=0,1,2,3 labels the transition (see
Fig.~\ref{fig:model}). For example, if the dot is in a zero-electron
ground state (GS0) and the electron tunnels in via the right lead,
the corresponding rate is $\Gamma_{\mathrm{R,0}}f_{\mathrm{R,0}}$,
as shown in Fig.~\ref{fig:model}. For $j$=0(1) the Fermi function is
$f_{i,0(1)}=1/\{\exp\left[(E_{\mathrm{B(A)}}-\mu_{\mathrm{i}})/k_{\mathrm{B}}T_{\mathrm{F}}\right]+1\}$
where $k_{\mathrm{B}}$ is the Boltzmann constant and
$T_{\mathrm{F}}$ is the temperature of the lead. For $j$=2,3 the
expression is analogous but the energy is lifted by the mutual
charging energy $e^{2}/C_{m}$.

In order to explain the experimental data presented above, we have
to take into account that the tunneling rates $\Gamma_{i,j}$ do
depend on the electronic wave function. They can be expressed as
$\Gamma_{i,j}=\gamma_{i}|c_{i,j}|^{2}$. The coefficients $c_{i,j}$
are the left ($i=L)$ and right ($i=R$) energy dependent components
of the eigenvector of the wave function corresponding to the bonding
($j$=0,3) and antibonding ($j$=1,2) states. The amplitudes
$\gamma_{i}$ are energy-independent parts of the $\Gamma_{i,j}$.

The rates describing the inter-dot processes, that is absorption
$\Gamma_{\mathrm{abs}}$ and emission $\Gamma_{\mathrm{em}}$ are
marked in Fig.~\ref{fig:model}. In the following, we assume that the
double quantum dot is coupled to a bosonic bath in thermal
equilibrium described by the Bose-Einstein distribution function
$n_{\mathrm{B}}(E,T_{\mathrm{B}})=1/\left[\exp\left(E/k_{\mathrm{B}}T_{\mathrm{B}}\right)-1\right]$.
The temperature $T_{\mathrm{B}}$ of this bath is determined by the
current of the QPC and the base temperature of the cryostat. In case
of coupling to acoustic phonons, the emission and absorption rates
can be expressed as (derivation is presented in Appendix A):
\begin{equation}\label{eq:FinalEmAbsRates}
\Gamma_{\mathrm{em/abs}}=\sum_{\lambda}\gamma_{\lambda}
\Delta^{n_{\lambda}}\left(n_{\mathrm{B}} \left (
\Delta,T_{\mathrm{B}} \right ) + \frac{1}{2} \mp \frac{1}{2}\right)
g_{\lambda}\left(\Delta\right),
\end{equation}
where the index $\lambda$ denotes piezoelectric transversal phonons
(pe,T), piezoelectric longitudinal phonons (pe,L) or longitudinal
deformation potential coupling phonons (dp,L). The exponent
$n{_\lambda}$ is 1 for piezoelectric phonons and $n_{\lambda}$=3 for
deformation potential coupling.\cite{Bockelmann,ThomasIhn} The upper
(lower) sign stands for emission (absorption) of energy quanta. The
values of the energy-independent coefficients $\gamma_{\lambda}$ are
given in Appendix~\ref{AppSec:AbsEmRates}. The form-factor
$g_{\lambda}(\Delta)$ is represented as:
\begin{equation}\label{eq:FunctionG}
g_{\lambda}\left(\Delta\right)=c_{\mathrm{L},\mathrm{B}}^{2}c_{\mathrm{L},\mathrm{A}}^{2}
\exp\left[-\left(\frac{\Delta r_0}{\hbar
c_{\lambda}}\right)^2\right] h_{\lambda}\left(\frac{\Delta d}{\hbar
c_{\lambda}}\right)
\end{equation}
where $r_0$ denotes the radius of a single QD, $d$ is the distance
between the dots, $c_{\lambda}$ is the speed of sound of
$\lambda$-phonons. The complete expression for the double-dot
geometry factor $h_{\lambda}\left(\Delta d/\hbar c_{\lambda}\right)$
is given in Appendix~\ref{AppSec:AbsEmRates}. In
Eq.~(\ref{eq:FunctionG}) the first factor
$c_{\mathrm{L},\mathrm{B}}^{2}c_{\mathrm{L},\mathrm{A}}^{2}$ is
related to the symmetry of the double quantum dot wave function. It
suppresses transitions for the asymmetric system. The second term of
Eq.~(\ref{eq:FunctionG}) refers to the shape of the individual dots.
It gives a high energy cut-off for phonon wavelengths much smaller
than the size of a single dot $r_0$. The last term of
Eq.~(\ref{eq:FunctionG}) arises from the separation of the single
dots in the double dot system. It suppresses small energy absorption
for phonon wavelengths much larger than $d$. Only phonons with a
wavelength comparable or larger than the DQD separation can interact
with the electron. For large energies this term has oscillatory
behavior\cite{Fujisawa:Science98}.

To investigate the influence of the QPCs on the dot presented in
Fig.~\ref{fig:detuning} in terms of rates and occupation
probabilities defined in Fig.~\ref{fig:model} we only take into
account the processes surrounded by the dashed line. This is
reasonable, due to a large mutual charging energy relative to the
tunneling coupling. Counting the electrons passing through the right
barrier leads to the following expression of the current through the
double dot (the derivation is presented in
Appendix~\ref{AppSec:RateEquation}):
\begin{equation}\label{eq:current0}
\begin{split}
I_{dot}
=&|e|\left[
    -p_{\mathrm{GS1}}\Gamma_{\mathrm{R2}}f_{\mathrm{R2}}-p_{\mathrm{EX1}}\Gamma_{\mathrm{R3}}f_{\mathrm{R3}}\right.\\
    &\left.+p_{\mathrm{GS2}}\Gamma_{\mathrm{R2}}\left(1-f_{\mathrm{R2}}\right)+p_{\mathrm{GS2}}\Gamma_{\mathrm{R3}}\left(1-f_{\mathrm{R3}}\right)
\right].
\end{split}
\end{equation}
The occupation probabilities $p_{\mathrm{GS1}}$, $p_{\mathrm{EX1}}$
and $p_{\mathrm{GS2}}$ are functions of $\Gamma_{i,j}$ and $f_{i,j}$
resulting from a stationary solution of the master equation (see
Appendix~\ref{AppSec:RateEquation}).

\section{Results and interpretation}\label{sec:Results}
We used expression~(\ref{eq:current0}) to fit the data shown in
Fig.~\ref{fig:detuning}. During the fitting procedure, the 8 traces
shown in Fig.~\ref{fig:detuning}(a) and (b) and an additional set of
16 traces being a combination of
$I_{\mathrm{QPC1,2}}=0,\pm50,\pm75,\pm100$ nA were fitted
simultaneously. The following fitting parameters were shared:
amplitudes $\gamma_{\mathrm{L}}$, $\gamma_{\mathrm{R}}$ and
tunneling coupling $t$. The only parameter specific for each trace
was the temperature of the phonon bath $T_{\mathrm{b}}$.

The fit represents the overall shape of the measured data very well.
The extracted tunneling coupling is $t=50~\mu$eV. In
Fig.~\ref{fig:SD}(a) the grey (green online) solid line is the
calculated boundary between the honeycomb cells assuming a tunneling
coupling of 50~$\mu$eV, the black (red online) crosses are the
maxima position of the DQD current peaks and the black dashed line
is the boundary assuming $t=0$. Unfortunately, the sample was not
stable enough to map a stability diagram with a resolution high
enough to determine the tunneling coupling directly. However, the
$t$ obtained from the fits seems to be reasonable and matches with
the data presented on the stability diagram.

The temperature of the phonon bath $T_{\mathrm{B}}$ obtained from
the fits varies from 0.6~K (blue trace in
Fig.~\ref{fig:detuning}(a)) to 1.2~K (red trace in
Fig.~\ref{fig:detuning}(b)). The difference between the temperatures
$T_\mathrm{b}$ and the electronic temperature $T_\mathrm{f}$=100~mK
gives rise to the DQD current.

The extracted amplitude $\gamma_{\mathrm{R}}$ is of the order of
7.8~MHz and $\gamma_{\mathrm{L}}$ is about 0.5~GHz. This is in
agreement with our previous statement that the right barrier is more
opaque than the left one. This leads to a deviation from perfect
antisymmetry of the dot current along the detuning line, i.e., to
suppressed dot current for negative detuning.

The tail of the curve at large detuning is mainly determined by the
amplitudes $c_{\mathrm{L},\mathrm{B}}^{2}
c_{\mathrm{L},\mathrm{A}}^{2}$ present in Eq.~(\ref{eq:FunctionG})
that drop to zero like $t^2/\delta^2$ and the Bose-Einstein
distribution function given in Eq.~(\ref{eq:FinalEmAbsRates}).
Although, the real distribution of the bosonic environment is not
necessarily equilibrated and could have another form, the
qualitative agreement with the data does not depend strongly on the
details of this distribution.

Using the parameters obtained in the fits, we calculated the current
as a function of the gate voltages using the model containing all
four charge states presented in Fig.~\ref{fig:model}. In the absence
of the QPC current ($T_{\mathrm{b}}=T_{f}$=70~mK), the result is
shown in Fig.~\ref{fig:SD}(c). The asymmetry and magnitude of the
current along the upper dashed line is reproduced very well. This is
not the case for the lower branch indicating transitions
$(\mathrm{M},\mathrm{N}) \leftrightarrow (\mathrm{M+1},\mathrm{N})$.
This may be due to a change of $\gamma_{\mathrm{L}}$ and
$\gamma_{\mathrm{R}}$, which we assumed to be constant in our model
but which may change in the experiment. For 1~mV DC bias across the
QPC2 (Fig.~\ref{fig:SD}(b)), the bosonic temperature is around
$T_{\mathrm{b}}$=0.8~K. The corresponding calculations are shown in
Fig.~\ref{fig:SD}(d). The red triangle indicates the region with the
QPC induced current that is in agreement with the measured data in
(b).

To investigate the dependence of the phonon temperature on the QPC
current we calculated $T_{\mathrm{b}}$ for every point from
Fig.~\ref{fig:addition}(a) using the values obtained in previous
fits. The results are shown in Fig.~\ref{fig:addition}(b). For small
QPC currents the error bars are large and no clear dependence is
visible. For QPC currents above 50 nA the dependence is quadratic
$T_{\mathrm{b}}\sim I_{\mathrm{QPC}}^2$, which means that the
temperature of the bosonic bath is proportional to the power emitted
by the QPC. The reconstructed temperatures corresponding to the
measurement presented in Fig.~\ref{fig:fits}(a) are plotted in
Fig.~\ref{fig:fits}(b). For very small DQD current it is impossible
to estimate the temperature of the phonon bath with sufficient
accuracy.

The saturation of the DQD current for large QPC currents cannot be
attributed to the high occupation probability of the antibonding
state $p_{\mathrm{EX1}}$. We have estimated that the
$p_{\mathrm{EX1}}$ value does not exceed a few percent and most of
the time the dot is occupied by one electron in its ground state.
The maximum current is determined by the right tunneling barrier.

A series of experiments reporting the observation of a DQD current
induced by a single and independently biased QPC, is described in
Refs.~\onlinecite{Khrapai:97.17,Khrapai:2008.1,Khrapai:2008.2}.
These experiments were performed in the regime of large DQD current,
strong bias voltage applied to the QPC and large tunneling coupling,
which is similar to our situation here. The DQD current was related
to inelastic relaxation of electrons in partly transmitting 1D
channels of the QPC~\cite{Khrapai:97.17} and qualitatively
consistent with an energy transfer mechanism based on nonequilibrium
acoustic phonons.~\cite{Khrapai:2008.1,Khrapai:2008.2} In contrast
to our experimental data, in these experiments the DQD current was
large when the conductance of the QPC was tuned to $e^{2}/h$ and
strongly suppressed in plateau regions.

Recent research has proven that the absorption of a photon can be
the dominant process~\cite{Gustavsson:99,gustavsson:152101} in
similar situations. However, these time-resolved experiments were
performed in a different regime, where the dominant tunneling rates
are of order of 1 kHz, whereas in our system the double quantum dot
is much more strongly coupled to the leads. Another difference is
the presence of a Ti top-gate in our structure, that screens the
electrostatic interaction between the DQD and the QPCs. Due to the
lower sensitivity of direct dot current measurements compared to the
time-resolved technique it is not possible to observe the gap in the
DQD current when $\left|eV_{QPC}\right|<\delta$. Calculations of the
emission and absorption rates in a DQD induced by electron-photon
interaction show, that the effect is irrelevant compared to the
emission and absorption of phonons discussed here.~\cite{ThomasIhn}

We have also tested the possibility that the entire dot current in
the region forbidden by the Coulomb blockade is due to a thermopower
effect induced by different temperatures in the source and the
drain. We found, that it would be only possible if the temperature
difference between source and drain lead was larger than 1K for a
QPC current of 100 nA, which is one order of magnitude larger than
expected.~\cite{Gustavsson:QPC} Another argument against a
thermopower model is that always the drain lead would have to be
warmer, even if the far QPC1 (that couples better to the drain lead)
was biased. Even so, the thermopower model did not describe the data
as well as the emission/absorption model.

\section{Conclusions}\label{sec:Conclusions}

We have presented the influence of two independent quantum point
contacts on a double quantum dot. A number of of questions arising
during the investigation and presented in Sec.~\ref{sec:Mechanisms}
could be answered.

In the first place, we established the possible dominant mechanism
of energy transfer between the double quantum dot and the quantum
point contact in the investigated regime. Driving current through
the QPC leads to emission of energy that increases the temperature
of the bosonic environment. We identify these bosons as acoustic
phonons. To model the interaction of the phonons with the double
quantum dot we have assumed that their energy distribution is
described by Bose-Einstein statistics.

Another important point is a non-additive effect of both QPC
currents. It is understood in terms of the temperature of the
bosonic bath. We find that the DQD current is proportional to the
power emitted by the QPCs.

Next, we interpreted the leveling off of the DQD current as a QPC
current is increased. For large QPC powers (above 0.1~nW) the
temperature of the phonon bath increases linearly. The observed
saturation of the current is due to the finite transparency of the
tunneling barriers and not to the high occupation probabilities
$p_{EX1}$.

The polarity dependence (Fig.~\ref{fig:addition}) cannot be
explained within the discussed model and its origin remains to be
investigated. It would be interesting to further investigate this
effect, for example, by using different geometrical arrangements.

Finally, we observed strong deviation from perfect antisymmetry of
the dot current along the detuning line when the QPC current is
driven. It can be attributed to the asymmetry of the source and
drain barriers.

All the measurements were performed with both QPCs tuned to their
first plateau. Thus we can exclude the influence of shot-noise
phenomena in the quantum point contacts.

\appendix\label{Appendix}

\section{Absorption and emission rates in a DQD induced by electron-phonon interaction}\label{AppSec:AbsEmRates}

Generally, the emission and absorption rates can be expressed using
Fermi's golden rule:
\begin{equation}\label{eq:FermiRule1}
\Gamma_{\mathrm{em/abs}}=\frac{2\pi}{\hbar}\sum_{\mathbf{q},\lambda}
\left | \left \langle
\Phi_{\mathrm{B}}|H_\mathrm{e-p}\left(\mathbf{r}\right)|
\Phi_{\mathrm{A}} \right \rangle \right |^{2} \delta \left (
\Delta-\hbar \omega_{\mathbf{q},\lambda}\right),
\end{equation}
where the sum extends over all wave vectors $\mathbf{q}$. The index
$\lambda$ denotes the type of acoustic phonons and their coupling in
GaAs: piezoelectric longitudinal (pe,L), piezoelectric transversal
(pe,T) and longitudinal, deformation potential coupling (dp,L). The
phonons have linear dispersion relation
$\omega_{\mathbf{q},\lambda}=c_{\lambda}|\mathbf{q}|$. $\Phi_{\mathrm{B}}$
and $\Phi_{\mathrm{A}}$ are wave functions of bonding and
antibonding states separated by energy $\Delta$. The interaction
hamiltonian $H_\mathrm{e-p}$ can be written as a sum of
piezoelectric interaction $H_{\mathrm{pe}}$ and deformation
potential coupling $H_{\mathrm{dp}}$:
\begin{equation}\label{eq:HamiltonianGeneral}
H_\mathrm{e-p}\left(\mathbf{r}\right)=H_{\mathrm{pe}}\left(\mathbf{r}\right)
+ H_{\mathrm{dp}}\left(\mathbf{r}\right),
\end{equation}
with
\begin{align}
   H_{\mathrm{pe}}\left(\mathbf{r}\right)&=-\frac{\left|e\right|d_{14}}{2\epsilon\epsilon_0\sqrt{NM}}
     \sum_{\mathbf{q},\lambda}\left(\frac{\hbar}{2\omega_{\mathbf{q},\lambda}}\right)^{1/2}F\left(\mathbf{q}\right)e^{i \mathbf{q}\mathbf{r}} \notag \\
   &\times \left( a_{\mathbf{q},\lambda}+a_{-\mathbf{q},\lambda}^\dag \right), \notag \\
   H_{\mathrm{dp}}\left(\mathbf{r}\right)&=-\frac{-iD}{\sqrt{NM}}
     \sum_{\mathbf{q}}\left(\frac{\hbar}{2\omega_{\mathbf{q},\lambda}}\right)^{1/2}\left|\mathbf{q}\right|e^{i \mathbf{q}\mathbf{r}} \notag \\
   &\times \left( a_{\mathbf{q},\lambda}+a_{-\mathbf{q},\lambda}^\dag \right).  \label{eq:HamiltonianPEDP}
\end{align}
In above equations $d_{14}$ is an element of piezoelectric tensor,
$\epsilon_0$ is the vacuum permittivity, $\epsilon$ is the
dielectric constant, $N$ is the number of atoms in the crystal, M is
the atomic mass and D denotes deformation potential coupling
constant. The dimensionless function $F\left(\mathbf{q}\right)$ has
form:
\begin{equation*}\label{eq:functionF}
 F\left(\mathbf{q}\right)=\frac{1}{\left|\mathbf{q}\right|^2}\sum_{ikl}\left|\varepsilon_{ikl}\right|
 \left(e_{\lambda,k}q_l+e_{\lambda,l}q_k\right)q_i,
\end{equation*}
where $\varepsilon_{ikl}$ is the Levi-Civita symbol and
$e_{\lambda,i}$ is the $i$-component of the eigenvector associated
with mode $\lambda$. Inserting Eq.~(\ref{eq:HamiltonianGeneral})
into Eq.~(\ref{eq:FermiRule1}) leads to the following expression:
\begin{equation}\label{eq:FermiRule2}
\begin{split}
 \Gamma_{\mathrm{em/abs}}&=\left(
 \gamma_{\mathrm{pe}}\Delta
 \left[
 g_{\mathrm{pe,T}}\left(\Delta\right)+
 g_{\mathrm{pe,L}}\left(\Delta\right)\right]+
 \gamma_{\mathrm{dp}}\Delta^3g_{\mathrm{dp}}\left(\Delta\right)
 \right)\\
 & \times \left(n(\Delta)+\frac{1}{2}\mp\frac{1}{2}\right),
\end{split}
\end{equation}
where the upper sign refers to phonon absorption and the lower to
phonon emission. The constants $\gamma_\mathrm{pe}$ and
$\gamma_\mathrm{dp}$ are given by:
\begin{equation*}\label{eq:gamma}
\begin{split}
&\gamma_{\mathrm{pe}}=\frac{\left|e\right|^2d_{14}^2}{2\pi
\left(2\epsilon\epsilon_0\right)^2\rho\hbar^2c_{\lambda}^3}
 =5\times10^{10} \mathrm{\ meV}^{-1}\mathrm{s}^{-1},\\
&\gamma_{\mathrm{dp}}=\frac{D^2}{2\pi\hbar^4c_{\lambda}^5\rho}
  =7.3\times 10^{11} \mathrm{\ meV}^{-3}\mathrm{s}^{-1},
\end{split}
\end{equation*}
where $\rho$ is a density of the GaAs crystal. The energy dependent
functions $g_{\mathrm{pe}}\left(\Delta\right)$ and
$g_{\mathrm{dp}}\left(\Delta\right)$ are defined as:
\begin{align}
    g_{\mathrm{pe}}\left(\Delta\right)&=
     \sum_{\lambda}\frac{\hbar^3c_{\lambda}^3}{4\pi\Delta^2}
     \int d^3qF_{\lambda}^2\left(\pm\mathbf{q}\right) \notag \\
   & \times
     \left | \left \langle \Phi_{\mathrm{B}}|e^{\pm i \mathbf{q}\mathbf{r}} | \Phi_{\mathrm{A}}\right \rangle \right|^{2}
     \delta \left(\Delta-\hbar \omega_{\mathbf{q},\lambda}\right),\notag \\
   g_{\mathrm{dp}}\left(\Delta\right)&=
     \frac{\hbar^5c_{\lambda}^5}{4\pi\Delta^4}
     \int d^3q\left|\mathbf{q}\right|^2 \notag \\
   & \times
     \left | \left \langle \Phi_{\mathrm{B}}|e^{\pm i \mathbf{q}\mathbf{r}} | \Phi_{\mathrm{A}}\right \rangle \right|^{2}
     \delta \left(\Delta-\hbar \omega_{\mathbf{q},\lambda}\right). \label{eq:gPEDP}
\end{align}

Assuming negligible overlap between the wave functions of the two
dots and taking a gaussian-shaped single-electron wave function, the
matrix element is found to be:
\begin{equation}\label{eq:MatrixElement}
 \left | \left \langle \Phi_{\mathrm{B}}|e^{\pm i \mathbf{q}\mathbf{r}} | \Phi_{\mathrm{A}}\right \rangle
 \right|^{2}=
 2c_{\mathrm{L},\mathrm{B}}^{2}c_{\mathrm{L},\mathrm{A}}^{2}e^{-\left(\mathbf{q}\mathbf{r_0}\right)^2}
 \left[1-\cos\left(\mathbf{q}\mathbf{d}\right)\right],
\end{equation}
where $d$ is the distance between the dots and $r_0$ is the radius
of a single dot. Inserting (\ref{eq:MatrixElement}) into
Eqs.~(\ref{eq:gPEDP}) gives:
\begin{equation}\label{eq:gLambda}
    g_{\mathrm{\lambda}}\left(\Delta\right)=
    c_{\mathrm{L},\mathrm{B}}^{2}c_{\mathrm{L},\mathrm{A}}^{2}
    e^{-\left(\mathbf{q}\mathbf{r_0}\right)^2}
    h_{\lambda}\left(\frac{\Delta d}{\hbar c_{\lambda}}\right)
\end{equation}
For piezoelectric transversal phonons the above expression was
calculated by averaging the function
$f_{\mathrm{pe,T}}\left({\mathbf{q}}\right)$ over all possible
transversal directions. The geometry factors
$h_{\lambda}\left(\eta=\Delta d/\hbar c_{\lambda}\right)$ are given
by:
\begin{align}
   h_{\mathrm{pe,L}}\left(\eta\right)&= \frac{24}{35}+72\eta^{-7}\left[9\eta\left(\eta^2-10\right)\cos\eta\right. \notag \\
   &+ \left.\left(\eta^4+39\eta^2+90\right)\sin\eta\right], \notag \\
   h_{\mathrm{pe,T}}\left(\eta\right)&= \frac{32}{35}+16\eta^{-7}\left[\eta\left(\eta^4-51\eta^2+405\right)\cos\eta\right. \notag \\
   &- \left.3\left(3\eta^4-62\eta^2+135\right)\sin\eta\right], \notag \\
   h_{\mathrm{dp,L}}\left(\eta\right)&= 2-2\eta^{-1}\sin\eta.  \label{eq:h}
\end{align}

Combining Eq.~(\ref{eq:h}) with Eq.~(\ref{eq:gLambda}) and inserting
the result into Eq.~(\ref{eq:FermiRule2}) gives a complete
expression for the absorption and emission rates.

\section{Rate equation}\label{AppSec:RateEquation}

To relate the DQD current to the tunneling rates we write down the
rate equation for the occupation of the states:
\begin{widetext}
\begin{equation}\label{eq:master}
\frac{d}{dt}\!\left[\!\begin{array}{c}
            \!p_{\mathrm{GS1}}\!\\
            \!p_{\mathrm{EX1}}\! \\
            \!p_{\mathrm{GS2}}\! \\
\end{array} \!\right]\!=\!
\left[\!\begin{array}{ccc}
     \!-\left(\!\Gamma_{\mathrm{abs}}\!+\!A_2\!\right)\!&\!\Gamma_{\mathrm{em}}\!&\!B_2\!\\
     \!\Gamma_{\mathrm{abs}}\!&\!-\!\left(\!\Gamma_{\mathrm{em}}\!+\!A_3\!\right)\!&\!B_3\!\\
     \!A_2\!&\!A_3\!&\!-\left(\!B_2\!+\!B_3\!\right)\\
\end{array}\!\right]\!
\left[\!\begin{array}{c}
     \!p_{\mathrm{GS1}}\! \\
     \!p_{\mathrm{EX1}}\! \\
     \!p_{\mathrm{GS2}}\! \\
\end{array}\!\right],
\end{equation}
\end{widetext}
with additional condition
$p_{\mathrm{GS1}}+p_{\mathrm{EX1}}+p_{\mathrm{GS2}}=1$. The terms
$A_j$ and $B_j$ are defined as:
\begin{equation}\label{eq:AB}
\begin{split}
&A_j=\sum_{i=\mathrm{L,R}}\Gamma_{i,j}f_{i,j}\\
&B_j=\sum_{i=\mathrm{L,R}}\Gamma_{i,j}\left(1-f_{i,j}\right).
\end{split}
\end{equation}
To find the expression for the current flowing through a DQD, we
take the right barrier as a current reference. It means that, if an
electron passes the right barrier to the left (right), its
contribution to the DQD current is positive (negative).
\begin{equation}\label{eq:current}
\begin{split}
I_{dot}=&|e|\left[
    -p_{\mathrm{GS1}}\Gamma_{\mathrm{R2}}f_{\mathrm{R2}}-p_{\mathrm{EX1}}\Gamma_{\mathrm{R3}}f_{\mathrm{R3}}\right.\\
    &\left. + p_{\mathrm{GS2}}\Gamma_{\mathrm{R2}}\left(1-f_{\mathrm{R2}}\right)+p_{\mathrm{GS2}}\Gamma_{\mathrm{R3}}\left(1-f_{\mathrm{R3}}\right)
\right],
\end{split}
\end{equation}
The first (second) term of Eq.~(\ref{eq:current}) corresponds to the
electrons moving from the bonding (antibonding) state to the right
lead and the third and fourth term to the electrons entering the
ground or excited state of the dot from the right lead. By inserting
a stationary solution of Eq.~(\ref{eq:master}) into
Eq.~(\ref{eq:current}) one obtains an expression for the steady
state DQD current.

\bibliographystyle{apsrev}

\begin{thebibliography}{32}
\expandafter\ifx\csname natexlab\endcsname\relax\def\natexlab#1{#1}\fi
\expandafter\ifx\csname bibnamefont\endcsname\relax
  \def\bibnamefont#1{#1}\fi
\expandafter\ifx\csname bibfnamefont\endcsname\relax
  \def\bibfnamefont#1{#1}\fi
\expandafter\ifx\csname citenamefont\endcsname\relax
  \def\citenamefont#1{#1}\fi
\expandafter\ifx\csname url\endcsname\relax
  \def\url#1{\texttt{#1}}\fi
\expandafter\ifx\csname urlprefix\endcsname\relax\def\urlprefix{URL }\fi
\providecommand{\bibinfo}[2]{#2}
\providecommand{\eprint}[2][]{\url{#2}}

\bibitem[{\citenamefont{Reed et~al.}(1989)\citenamefont{Reed, Randall,
  Luscombe, Frensley, Aggarwal, Matyi, Moore, and Wetsel}}]{Reed}
\bibinfo{author}{\bibfnamefont{M.}~\bibnamefont{Reed}},
  \bibinfo{author}{\bibfnamefont{J.}~\bibnamefont{Randall}},
  \bibinfo{author}{\bibfnamefont{J.}~\bibnamefont{Luscombe}},
  \bibinfo{author}{\bibfnamefont{W.}~\bibnamefont{Frensley}},
  \bibinfo{author}{\bibfnamefont{R.}~\bibnamefont{Aggarwal}},
  \bibinfo{author}{\bibfnamefont{R.}~\bibnamefont{Matyi}},
  \bibinfo{author}{\bibfnamefont{T.}~\bibnamefont{Moore}}, \bibnamefont{and}
  \bibinfo{author}{\bibfnamefont{A.}~\bibnamefont{Wetsel}}, in
  \emph{\bibinfo{booktitle}{Festk{\"{o}}rperprobleme 29}} (\bibinfo{publisher}{Springer Berlin/Heidelberg},
  \bibinfo{year}{1989}),  pp. \bibinfo{pages}{267--283}.

\bibitem[{\citenamefont{van~der Wiel et~al.}(2002)\citenamefont{van~der Wiel,
  De~Franceschi, Elzerman, Fujisawa, Tarucha, and Kouwenhoven}}]{vanderWiel:1}
\bibinfo{author}{\bibfnamefont{W.~G.} \bibnamefont{van~der Wiel}},
  \bibinfo{author}{\bibfnamefont{S.}~\bibnamefont{De~Franceschi}},
  \bibinfo{author}{\bibfnamefont{J.~M.} \bibnamefont{Elzerman}},
  \bibinfo{author}{\bibfnamefont{T.}~\bibnamefont{Fujisawa}},
  \bibinfo{author}{\bibfnamefont{S.}~\bibnamefont{Tarucha}}, \bibnamefont{and}
  \bibinfo{author}{\bibfnamefont{L.~P.} \bibnamefont{Kouwenhoven}},
  \bibinfo{journal}{Rev. Mod. Phys.} \textbf{\bibinfo{volume}{75}},
  \bibinfo{pages}{1} (\bibinfo{year}{2002}).

\bibitem[{\citenamefont{Hanson et~al.}(2007)\citenamefont{Hanson, Kouwenhoven,
  Petta, Tarucha, and Vandersypen}}]{Hanson}
\bibinfo{author}{\bibfnamefont{R.}~\bibnamefont{Hanson}},
  \bibinfo{author}{\bibfnamefont{L.~P.} \bibnamefont{Kouwenhoven}},
  \bibinfo{author}{\bibfnamefont{J.~R.} \bibnamefont{Petta}},
  \bibinfo{author}{\bibfnamefont{S.}~\bibnamefont{Tarucha}}, \bibnamefont{and}
  \bibinfo{author}{\bibfnamefont{L.~M.~K.} \bibnamefont{Vandersypen}},
  \bibinfo{journal}{Rev. Mod. Phys.} \textbf{\bibinfo{volume}{79}},
  \bibinfo{pages}{1217} (\bibinfo{year}{2007}).

\bibitem[{\citenamefont{Fujisawa and
  Tarucha}(1997{\natexlab{a}})}]{Fujisawa:36}
  \bibinfo{author}{\bibfnamefont{T.}~\bibnamefont{Fujisawa}} \bibnamefont{and}
  \bibinfo{author}{\bibfnamefont{S.}~\bibnamefont{Tarucha}}, \bibinfo{journal}{Jpn. J. of App.
  Phys.} \textbf{\bibinfo{volume}{36}}, \bibinfo{pages}{4000}
  (\bibinfo{year}{1997}{\natexlab{a}}).

\bibitem[{\citenamefont{Fujisawa and
  Tarucha}(1997{\natexlab{b}})}]{Fujisawa:21:247}
\bibinfo{author}{\bibfnamefont{T.}~\bibnamefont{Fujisawa}} \bibnamefont{and}
  \bibinfo{author}{\bibfnamefont{S.}~\bibnamefont{Tarucha}},
  \bibinfo{journal}{Superlattices and Microstructures}
  \textbf{\bibinfo{volume}{21}}, \bibinfo{pages}{247}
  (\bibinfo{year}{1997}{\natexlab{b}}).

\bibitem[{\citenamefont{Oosterkamp et~al.}(1997)\citenamefont{Oosterkamp,
  Kouwenhoven, Koolen, van~der Vaart, and Harmans}}]{Oosterkamp}
\bibinfo{author}{\bibfnamefont{T.~H.} \bibnamefont{Oosterkamp}},
  \bibinfo{author}{\bibfnamefont{L.~P.} \bibnamefont{Kouwenhoven}},
  \bibinfo{author}{\bibfnamefont{A.~E.~A.} \bibnamefont{Koolen}},
  \bibinfo{author}{\bibfnamefont{N.~C.} \bibnamefont{van~der Vaart}},
  \bibnamefont{and} \bibinfo{author}{\bibfnamefont{C.~J. P.~M.}
  \bibnamefont{Harmans}}, \bibinfo{journal}{Phys. Rev. Lett.}
  \textbf{\bibinfo{volume}{78}}, \bibinfo{pages}{1536} (\bibinfo{year}{1997}).

\bibitem[{\citenamefont{Oosterkamp et~al.}(1998)\citenamefont{Oosterkamp,
  Fujisawa, van~der Wiel, Ishibashi, Hijman, Tarucha, and
  Kouwenhoven}}]{Oosterkamp:Nature}
\bibinfo{author}{\bibfnamefont{T.~H.} \bibnamefont{Oosterkamp}},
  \bibinfo{author}{\bibfnamefont{T.}~\bibnamefont{Fujisawa}},
  \bibinfo{author}{\bibfnamefont{W.~G.} \bibnamefont{van~der Wiel}},
  \bibinfo{author}{\bibfnamefont{K.}~\bibnamefont{Ishibashi}},
  \bibinfo{author}{\bibfnamefont{R.~V.} \bibnamefont{Hijman}},
  \bibinfo{author}{\bibfnamefont{S.}~\bibnamefont{Tarucha}}, \bibnamefont{and}
  \bibinfo{author}{\bibfnamefont{L.~P.} \bibnamefont{Kouwenhoven}},
  \bibinfo{journal}{Nature} \textbf{\bibinfo{volume}{395}},
  \bibinfo{pages}{873} (\bibinfo{year}{1998}).

\bibitem[{\citenamefont{van~der Wiel et~al.}(1999)\citenamefont{van~der Wiel,
  Fujisawa, Oosterkamp, and Kouwenhoven}}]{vanderWiel:272}
\bibinfo{author}{\bibfnamefont{W.~G.} \bibnamefont{van~der Wiel}},
  \bibinfo{author}{\bibfnamefont{T.}~\bibnamefont{Fujisawa}},
  \bibinfo{author}{\bibfnamefont{T.~H.} \bibnamefont{Oosterkamp}},
  \bibnamefont{and} \bibinfo{author}{\bibfnamefont{L.~P.}
  \bibnamefont{Kouwenhoven}}, \bibinfo{journal}{Physica B}
  \textbf{\bibinfo{volume}{272}}, \bibinfo{pages}{31} (\bibinfo{year}{1999}).

\bibitem[{\citenamefont{Qin et~al.}(2001)\citenamefont{Qin, Holleitner, Eberl,
  and Blick}}]{Blick:PAT}
\bibinfo{author}{\bibfnamefont{H.}~\bibnamefont{Qin}},
  \bibinfo{author}{\bibfnamefont{A.~W.} \bibnamefont{Holleitner}},
  \bibinfo{author}{\bibfnamefont{K.}~\bibnamefont{Eberl}}, \bibnamefont{and}
  \bibinfo{author}{\bibfnamefont{R.~H.} \bibnamefont{Blick}},
  \bibinfo{journal}{Phys. Rev. B} \textbf{\bibinfo{volume}{64}},
  \bibinfo{pages}{241302 (R)} (\bibinfo{year}{2001}).

\bibitem[{\citenamefont{Switkes et~al.}(1999)\citenamefont{Switkes, Marcus,
  Campman, and Gossard}}]{Switkes}
\bibinfo{author}{\bibfnamefont{M.}~\bibnamefont{Switkes}},
  \bibinfo{author}{\bibfnamefont{C.~M.} \bibnamefont{Marcus}},
  \bibinfo{author}{\bibfnamefont{K.}~\bibnamefont{Campman}}, \bibnamefont{and}
  \bibinfo{author}{\bibfnamefont{A.~C.} \bibnamefont{Gossard}},
  \bibinfo{journal}{Science} \textbf{\bibinfo{volume}{283}},
  \bibinfo{pages}{1905} (\bibinfo{year}{1999}).

\bibitem[{\citenamefont{Platero and Aguado}(2004)}]{Platero}
\bibinfo{author}{\bibfnamefont{G.}~\bibnamefont{Platero}} \bibnamefont{and}
  \bibinfo{author}{\bibfnamefont{R.}~\bibnamefont{Aguado}},
  \bibinfo{journal}{Physics Reports} \textbf{\bibinfo{volume}{395}},
  \bibinfo{pages}{1} (\bibinfo{year}{2004}).

\bibitem[{\citenamefont{Field et~al.}(1993)\citenamefont{Field, Smith, Pepper,
  Ritchie, Frost, Jones, and Hasko}}]{Field:70.1311}
\bibinfo{author}{\bibfnamefont{M.}~\bibnamefont{Field}},
  \bibinfo{author}{\bibfnamefont{C.~G.} \bibnamefont{Smith}},
  \bibinfo{author}{\bibfnamefont{M.}~\bibnamefont{Pepper}},
  \bibinfo{author}{\bibfnamefont{D.~A.} \bibnamefont{Ritchie}},
  \bibinfo{author}{\bibfnamefont{J.~E.~F.} \bibnamefont{Frost}},
  \bibinfo{author}{\bibfnamefont{G.~A.~C.} \bibnamefont{Jones}},
  \bibnamefont{and} \bibinfo{author}{\bibfnamefont{D.~G.} \bibnamefont{Hasko}},
  \bibinfo{journal}{Phys. Rev. Lett.} \textbf{\bibinfo{volume}{70}},
  \bibinfo{pages}{1311} (\bibinfo{year}{1993}).

\bibitem[{\citenamefont{Elzerman et~al.}(2003)\citenamefont{Elzerman, Hanson,
  Greidanus, Willems~van Beveren, De~Franceschi, Vandersypen, Tarucha, and
  Kouwenhoven}}]{Elzerman2003}
\bibinfo{author}{\bibfnamefont{J.~M.} \bibnamefont{Elzerman}},
  \bibinfo{author}{\bibfnamefont{R.}~\bibnamefont{Hanson}},
  \bibinfo{author}{\bibfnamefont{J.~S.} \bibnamefont{Greidanus}},
  \bibinfo{author}{\bibfnamefont{L.~H.} \bibnamefont{Willems~van Beveren}},
  \bibinfo{author}{\bibfnamefont{S.}~\bibnamefont{De~Franceschi}},
  \bibinfo{author}{\bibfnamefont{L.~M.~K.} \bibnamefont{Vandersypen}},
  \bibinfo{author}{\bibfnamefont{S.}~\bibnamefont{Tarucha}}, \bibnamefont{and}
  \bibinfo{author}{\bibfnamefont{L.~P.} \bibnamefont{Kouwenhoven}},
  \bibinfo{journal}{Phys. Rev. B} \textbf{\bibinfo{volume}{67}},
  \bibinfo{pages}{161308 (R) } (\bibinfo{year}{2003}).

\bibitem[{\citenamefont{DiCarlo et~al.}(2004)\citenamefont{DiCarlo, Lynch,
  Johnson, Childress, Crockett, Marcus, Hanson, and Gossard}}]{DiCarlo}
\bibinfo{author}{\bibfnamefont{L.}~\bibnamefont{DiCarlo}},
  \bibinfo{author}{\bibfnamefont{H.~J.} \bibnamefont{Lynch}},
  \bibinfo{author}{\bibfnamefont{A.~C.} \bibnamefont{Johnson}},
  \bibinfo{author}{\bibfnamefont{L.~I.} \bibnamefont{Childress}},
  \bibinfo{author}{\bibfnamefont{K.}~\bibnamefont{Crockett}},
  \bibinfo{author}{\bibfnamefont{C.~M.} \bibnamefont{Marcus}},
  \bibinfo{author}{\bibfnamefont{M.~P.} \bibnamefont{Hanson}},
  \bibnamefont{and} \bibinfo{author}{\bibfnamefont{A.~C.}
  \bibnamefont{Gossard}}, \bibinfo{journal}{Phys. Rev. Lett.}
  \textbf{\bibinfo{volume}{92}}, \bibinfo{pages}{226801}
  (\bibinfo{year}{2004}).

\bibitem[{\citenamefont{Khrapai et~al.}(2006)\citenamefont{Khrapai, Ludwig,
  Kotthaus, Tranitz, and Wegscheider}}]{Khrapai:97.17}
\bibinfo{author}{\bibfnamefont{V.~S.} \bibnamefont{Khrapai}},
  \bibinfo{author}{\bibfnamefont{S.}~\bibnamefont{Ludwig}},
  \bibinfo{author}{\bibfnamefont{J.~P.} \bibnamefont{Kotthaus}},
  \bibinfo{author}{\bibfnamefont{H.~P.} \bibnamefont{Tranitz}},
  \bibnamefont{and}
  \bibinfo{author}{\bibfnamefont{W.}~\bibnamefont{Wegscheider}},
  \bibinfo{journal}{Phys. Rev. Lett.} \textbf{\bibinfo{volume}{97}},
  \bibinfo{pages}{176803} (\bibinfo{year}{2006}).

\bibitem[{\citenamefont{Khrapai
  et~al.}(2008{\natexlab{a}})\citenamefont{Khrapai, Ludwig, Kotthaus, Tranitz,
  and Wegscheider}}]{Khrapai:2008.1}
\bibinfo{author}{\bibfnamefont{V.~S.} \bibnamefont{Khrapai}},
  \bibinfo{author}{\bibfnamefont{S.}~\bibnamefont{Ludwig}},
  \bibinfo{author}{\bibfnamefont{J.~P.} \bibnamefont{Kotthaus}},
  \bibinfo{author}{\bibfnamefont{H.~P.} \bibnamefont{Tranitz}},
  \bibnamefont{and}
  \bibinfo{author}{\bibfnamefont{W.}~\bibnamefont{Wegscheider}},
  \bibinfo{journal}{Physica E: Low-dimensional Systems and Nanostructures}
  \textbf{\bibinfo{volume}{40}}, \bibinfo{pages}{995}
  (\bibinfo{year}{2008}{\natexlab{a}}).

\bibitem[{\citenamefont{Khrapai
  et~al.}(2008{\natexlab{b}})\citenamefont{Khrapai, Ludwig, Kotthaus, Tranitz,
  and Wegscheider}}]{Khrapai:2008.2}
\bibinfo{author}{\bibfnamefont{V.~S.} \bibnamefont{Khrapai}},
  \bibinfo{author}{\bibfnamefont{S.}~\bibnamefont{Ludwig}},
  \bibinfo{author}{\bibfnamefont{J.~P.} \bibnamefont{Kotthaus}},
  \bibinfo{author}{\bibfnamefont{H.~P.} \bibnamefont{Tranitz}},
  \bibnamefont{and}
  \bibinfo{author}{\bibfnamefont{W.}~\bibnamefont{Wegscheider}}
  (\bibinfo{year}{2008}{\natexlab{b}}), \bibinfo{note}{arXiv:0805.0724v1}.

\bibitem[{\citenamefont{Gustavsson et~al.}(2007)\citenamefont{Gustavsson,
  Studer, Leturcq, Ihn, Ensslin, Driscoll, and Gossard}}]{Gustavsson:99}
\bibinfo{author}{\bibfnamefont{S.}~\bibnamefont{Gustavsson}},
  \bibinfo{author}{\bibfnamefont{M.}~\bibnamefont{Studer}},
  \bibinfo{author}{\bibfnamefont{R.}~\bibnamefont{Leturcq}},
  \bibinfo{author}{\bibfnamefont{T.}~\bibnamefont{Ihn}},
  \bibinfo{author}{\bibfnamefont{K.}~\bibnamefont{Ensslin}},
  \bibinfo{author}{\bibfnamefont{D.~C.} \bibnamefont{Driscoll}},
  \bibnamefont{and} \bibinfo{author}{\bibfnamefont{A.~C.}
  \bibnamefont{Gossard}}, \bibinfo{journal}{Phys. Rev. Lett.}
  \textbf{\bibinfo{volume}{99}}, \bibinfo{pages}{206804}
  (\bibinfo{year}{2007}).

\bibitem[{\citenamefont{Fujisawa et~al.}(2006)\citenamefont{Fujisawa, Hayashi,
  Jung, Jeong, and Hirayama}}]{Fujisawa:2006:Single}
\bibinfo{author}{\bibfnamefont{T.}~\bibnamefont{Fujisawa}},
  \bibinfo{author}{\bibfnamefont{T.}~\bibnamefont{Hayashi}},
  \bibinfo{author}{\bibfnamefont{S.}~\bibnamefont{Jung}},
  \bibinfo{author}{\bibfnamefont{Y.-H.} \bibnamefont{Jeong}}, \bibnamefont{and}
  \bibinfo{author}{\bibfnamefont{Y.}~\bibnamefont{Hirayama}}, in
  \emph{\bibinfo{booktitle}{Quantum Computing in Solid State Systems}}
  (\bibinfo{publisher}{Springer},
  \bibinfo{year}{2006}),  pp. \bibinfo{pages}{279--287}.


\bibitem[{\citenamefont{Fujisawa et~al.}(1998)\citenamefont{Fujisawa,
  Oosterkamp, van~der Wiel, Broer, Aguado, Tarucha, and
  Kouwenhoven}}]{Fujisawa:Science98}
\bibinfo{author}{\bibfnamefont{T.}~\bibnamefont{Fujisawa}},
  \bibinfo{author}{\bibfnamefont{T.~H.} \bibnamefont{Oosterkamp}},
  \bibinfo{author}{\bibfnamefont{W.~G.} \bibnamefont{van~der Wiel}},
  \bibinfo{author}{\bibfnamefont{B.~W.} \bibnamefont{Broer}},
  \bibinfo{author}{\bibfnamefont{R.}~\bibnamefont{Aguado}},
  \bibinfo{author}{\bibfnamefont{S.}~\bibnamefont{Tarucha}}, \bibnamefont{and}
  \bibinfo{author}{\bibfnamefont{L.~P.} \bibnamefont{Kouwenhoven}},
  \bibinfo{journal}{Science} \textbf{\bibinfo{volume}{282}},
  \bibinfo{pages}{932} (\bibinfo{year}{1998}).

\bibitem[{\citenamefont{Onac et~al.}(2006)\citenamefont{Onac, Balestro, Willems~van
  Beveren, Hartmann, Nazarov, and Kouwenhoven}}]{Onac}
\bibinfo{author}{\bibfnamefont{E.}~\bibnamefont{Onac}},
  \bibinfo{author}{\bibfnamefont{F.}~\bibnamefont{Balestro}},
  \bibinfo{author}{\bibfnamefont{L.~H.} \bibnamefont{Willems~van Beveren}},
  \bibinfo{author}{\bibfnamefont{U.}~\bibnamefont{Hartmann}},
  \bibinfo{author}{\bibfnamefont{Y.~V.} \bibnamefont{Nazarov}},
  \bibnamefont{and} \bibinfo{author}{\bibfnamefont{L.~P.}
  \bibnamefont{Kouwenhoven}}, \bibinfo{journal}{Phys. Rev. Lett.}
  \textbf{\bibinfo{volume}{96}}, \bibinfo{pages}{176601}
  (\bibinfo{year}{2006}).

\bibitem[{\citenamefont{Zakka-Bajjani et~al.}(2007)\citenamefont{Zakka-Bajjani,
  Segala, Portier, Roche, Glattli, Cavanna, and Jin}}]{Zakka}
\bibinfo{author}{\bibfnamefont{E.}~\bibnamefont{Zakka-Bajjani}},
  \bibinfo{author}{\bibfnamefont{J.}~\bibnamefont{Segala}},
  \bibinfo{author}{\bibfnamefont{F.}~\bibnamefont{Portier}},
  \bibinfo{author}{\bibfnamefont{P.}~\bibnamefont{Roche}},
  \bibinfo{author}{\bibfnamefont{D.~C.} \bibnamefont{Glattli}},
  \bibinfo{author}{\bibfnamefont{A.}~\bibnamefont{Cavanna}}, \bibnamefont{and}
  \bibinfo{author}{\bibfnamefont{Y.}~\bibnamefont{Jin}},
  \bibinfo{journal}{Phys. Rev. Lett.} \textbf{\bibinfo{volume}{99}},
  \bibinfo{pages}{236803} (\bibinfo{year}{2007}).

\bibitem[{\citenamefont{Sigrist et~al.}(2004)\citenamefont{Sigrist, Fuhrer,
  Ihn, Ensslin, Driscoll, and Gossard}}]{sigrist:3558}
\bibinfo{author}{\bibfnamefont{M.}~\bibnamefont{Sigrist}},
  \bibinfo{author}{\bibfnamefont{A.}~\bibnamefont{Fuhrer}},
  \bibinfo{author}{\bibfnamefont{T.}~\bibnamefont{Ihn}},
  \bibinfo{author}{\bibfnamefont{K.}~\bibnamefont{Ensslin}},
  \bibinfo{author}{\bibfnamefont{D.~C.} \bibnamefont{Driscoll}},
  \bibnamefont{and} \bibinfo{author}{\bibfnamefont{A.~C.}
  \bibnamefont{Gossard}}, \bibinfo{journal}{Appl. Phys. Lett.}
  \textbf{\bibinfo{volume}{85}}, \bibinfo{pages}{3558} (\bibinfo{year}{2004}).

\bibitem[{\citenamefont{Fuhrer et~al.}(2002)\citenamefont{Fuhrer, Dorn,
  Luscher, Heinzel, Ensslin, Wegscheider, and Bichler}}]{Fuhrer}
\bibinfo{author}{\bibfnamefont{A.}~\bibnamefont{Fuhrer}},
  \bibinfo{author}{\bibfnamefont{A.}~\bibnamefont{Dorn}},
  \bibinfo{author}{\bibfnamefont{S.}~\bibnamefont{Luscher}},
  \bibinfo{author}{\bibfnamefont{T.}~\bibnamefont{Heinzel}},
  \bibinfo{author}{\bibfnamefont{K.}~\bibnamefont{Ensslin}},
  \bibinfo{author}{\bibfnamefont{W.}~\bibnamefont{Wegscheider}},
  \bibnamefont{and} \bibinfo{author}{\bibfnamefont{M.}~\bibnamefont{Bichler}},
  \bibinfo{journal}{Superlattices and Microstructures}
  \textbf{\bibinfo{volume}{31}}, \bibinfo{pages}{19} (\bibinfo{year}{2002}).

\bibitem[{\citenamefont{Davies et~al.}(1995)\citenamefont{Davies, Larkin, and
  Sukhorukov}}]{Davies:1}
\bibinfo{author}{\bibfnamefont{J.~H.} \bibnamefont{Davies}},
  \bibinfo{author}{\bibfnamefont{I.~A.} \bibnamefont{Larkin}},
  \bibnamefont{and} \bibinfo{author}{\bibfnamefont{E.~V.}
  \bibnamefont{Sukhorukov}}, \bibinfo{journal}{J. of App. Phys.}
  \textbf{\bibinfo{volume}{77}}, \bibinfo{pages}{4504} (\bibinfo{year}{1995}).

\bibitem[{pot()}]{potential}
\bibinfo{note}{In order to estimate the electrostatic potential in a 2DEG plane
  created by the AFM-defined oxide lines and titanium top-gates, we treated the
  oxide lines as patterned metalic gates covering the surface of the
  semiconductor heterostructure. The Ti top-gates were definied in the same
  manner. In our calculations the voltage applied to the gates corresponding to
  the oxide lines was 10 times larger than the voltage applied to the Ti
  top-gates.}

\bibitem[{\citenamefont{Gustavsson et~al.}(2008)\citenamefont{Gustavsson,
  Shorubalko, Leturcq, Sch\"{o}n, and Ensslin}}]{gustavsson:152101}
\bibinfo{author}{\bibfnamefont{S.}~\bibnamefont{Gustavsson}},
  \bibinfo{author}{\bibfnamefont{I.}~\bibnamefont{Shorubalko}},
  \bibinfo{author}{\bibfnamefont{R.}~\bibnamefont{Leturcq}},
  \bibinfo{author}{\bibfnamefont{S.}~\bibnamefont{Sch\"{o}n}},
  \bibnamefont{and} \bibinfo{author}{\bibfnamefont{K.}~\bibnamefont{Ensslin}},
  \bibinfo{journal}{App. Phys. Lett.} \textbf{\bibinfo{volume}{92}},
  \bibinfo{pages}{152101} (\bibinfo{year}{2008}).

\bibitem[{\citenamefont{Bockelmann}(1994)}]{Bockelmann}
\bibinfo{author}{\bibfnamefont{U.}~\bibnamefont{Bockelmann}},
  \bibinfo{journal}{Phys. Rev. B} \textbf{\bibinfo{volume}{50}},
  \bibinfo{pages}{17271} (\bibinfo{year}{1994}).

\bibitem[{\citenamefont{Blick et~al.}(1998)\citenamefont{Blick, Pfannkuche,Haug, Klitzing, and Eberl}}]{Blick:molecular}
\bibinfo{author}{\bibfnamefont{R.~H.} \bibnamefont{Blick}},
  \bibinfo{author}{\bibfnamefont{D.}~\bibnamefont{Pfannkuche}},
  \bibinfo{author}{\bibfnamefont{R.~J.} \bibnamefont{Haug}},
  \bibinfo{author}{\bibfnamefont{K.~v.} \bibnamefont{Klitzing}},
  \bibnamefont{and} \bibinfo{author}{\bibfnamefont{K.}~\bibnamefont{Eberl}},
  \bibinfo{journal}{Phys. Rev. Lett.} \textbf{\bibinfo{volume}{80}},
  \bibinfo{pages}{4032} (\bibinfo{year}{1998}).

\bibitem[{\citenamefont{Ihn}()}]{ThomasIhn}
\bibinfo{author}{\bibfnamefont{T.}~\bibnamefont{Ihn}},
  \bibinfo{note}{unpublished}.

\bibitem[{\citenamefont{Gustavsson}(2008)}]{Gustavsson:QPC}
\bibinfo{author}{\bibfnamefont{S.}~\bibnamefont{Gustavsson}}, Ph.D. thesis,
  \bibinfo{school}{ETH Zurich}, \bibinfo{address}{Zurich, Switzerland}
  (\bibinfo{year}{2008}).

\end{thebibliography}

\end{document}